# Quantum and classical correlations in the one-dimensional $XY$ model with Dzyaloshinskii-Moriya interaction

Ben-Qiong Liu, [1] Bin Shao, [1*] Jun-Gang Li, [1] Jian Zou, [1] and Lian-Ao Wu[2]

[1] Key Laboratory of Cluster Science of Ministry of Education, and Department of Physics, Beijing Institute of Technology, Beijing 100081, China

[2] Department of Theoretical Physics and History of Science, The Basque Country University (EHU/UPV), Post Office Box 644, ES-48080 Bilbao, Spain and IKERBASQUE, Basque Foundation for Science, ES-48011 Bilbao, Spain

**Abstract**

We study the effect of Dzyaloshinskii-Moriya (DM) interaction on pairwise quantum discord, entanglement, and classical correlation in the anisotropic $XY$ spin-half chain. Analytical expressions for both quantum and classical correlations are obtained from the spin-spin correlation functions. We show that these pairwise quantities exhibit various behaviors in relation to the relative strengths of the DM interaction, the anisotropy and the magnetic intensity. We observe non-analyticities of the derivatives of both quantum and classical correlations with respect to the magnetic intensity at the critical point, with consideration of the DM interaction.

*

---

* Corresponding author: sbin610@bit.edu.cn

## I. INTRODUCTION

Entanglement, at the heart of quantum reality, has received great attention in various branches of quantum physics [1, 2] mostly because of its promising features exhibited in the quantum information processing. While the interest remains strong, recent researches have explored non-classical correlations other than entanglement, which may be employed as alternative resources for quantum technology [3-6]. Among the quantum correlations, the quantum discord introduced by Ollivier and Zurek [4] has been studied relatively comprehensively and is supposed to characterize all the nonclassical correlations in a bipartite state, including entanglement. While in pure states it coincides with the entanglement entropy, the quantum discord does not vanish in mixed separable states, even if entanglement is absent. This unique feature suggests that quantum correlations, in particular quantum discord, may have significant applications in revealing the advantage of certain quantum tasks and might be a more comprehensive resource than entanglement [7-21], as evidenced, e.g., by their behaviors in quantum critical phenomena. It was interesting to note that the quantum discord, in contrast to the entanglement, is able to signal quantum phase transitions [22, 23]. Moreover, study on the pairwise quantum discord in the thermodynamic limit of the $XY$ chain shows that the quantum discord for spin pairs farther than second neighbors could still signal a quantum phase transition (QPT), while the corresponding pairwise entanglement vanishes [24].

This paper studies behaviors of both quantum and classical correlations in the anisotropic $XY$ spin-half chain with the Dzyaloshinskii-Moriya (DM) interaction [25] $\sum_{\langle ij \rangle} \vec{D}_{ij} \cdot \left( \vec{S}_i \times \vec{S}_j \right)$, which arises from the spin-orbit coupling. Despite being small, this interaction can generate interesting effects [26, 27] and is crucial to the description of many antiferromagnetic systems, such as $Cu(C_6D_5COO)_2 3D_2O$ [28], $Yb_4As_3$ [29], $BaCu_2Si_2O_7$ [30], and $K_2V_3O_8$ [31]. It also plays a significant role in quantum dots [32] and in performing universal quantum computation [33, 34]. The behaviors of entanglement in spin chains [35, 36] with DM interaction were studied extensively. The influence of DM interaction on quantum phase interference of spins was discussed [37]. The entanglement transfer in a spin chain with DM interaction was examined [38, 39].

This paper is arranged as follows. Section II introduces the anisotropic $XY$ spin chain with DM interaction and describes briefly the techniques of correlation functions used to obtain our results. We analyze the behaviors of quantum and classical correlations under the influence of the DM interaction in Section III. We conclude our work in Section IV.

**II. QUANTUM AND CLASSICAL CORRELATIONS IN THE $XY$ CHAIN WITH DM INTERACTION**

Consider the anisotropic $XY$ spin-half chain with DM interaction in the $z$ direction,

$$H = \sum_{j=1}^{N}\left\{J\left[(1+\gamma)S_j^x S_{j+1}^x + (1-\gamma)S_j^y S_{j+1}^y + D\left(S_j^x S_{j+1}^y - S_j^y S_{j+1}^x\right)\right] - S_j^z\right\}, \quad (1)$$

where $S_j^\alpha$ ($\alpha = x, y, z$) are the spin-half operators at the $j$th lattice site, $N$ is the total number of spins, and the periodic boundary conditions ($S_{N+1}^\alpha = S_1^\alpha$) are used. The parameter $\gamma$ characterizes the degree of anisotropy, $D$ denotes the intensity of the DM interaction along the $z$ direction and $J$ is the strength of the inverse of the external transverse magnetic field.

The two-spin density matrix $\rho_{ij}$ for this system in the computational basis ($|\uparrow\uparrow\rangle$, $|\uparrow\downarrow\rangle$, $|\downarrow\uparrow\rangle$, $|\downarrow\downarrow\rangle$) is given by [40]

$$\rho_{ij} = \begin{pmatrix} u_{ij}^+ & 0 & 0 & y_{ij} \\ 0 & \omega_{ij}^+ & x_{ij} & 0 \\ 0 & x_{ij} & \omega_{ij}^- & 0 \\ y_{ij} & 0 & 0 & u_{ij}^- \end{pmatrix}, \quad (2)$$

where matrix elements can be written in terms of one- and two-point correlation functions,

$$u_{ij}^{\pm} = \frac{1}{4} \pm \frac{\langle S_i^z \rangle}{2} \pm \frac{\langle S_j^z \rangle}{2} + \langle S_i^z S_j^z \rangle, \tag{3}$$

$$\omega_{ij}^{\pm} = \frac{1}{4} \mp \frac{\langle S_i^z \rangle}{2} \pm \frac{\langle S_j^z \rangle}{2} - \langle S_i^z S_j^z \rangle, \tag{4}$$

$$x_{ij} = \langle S_i^x S_j^x \rangle + \langle S_i^y S_j^y \rangle, \tag{5}$$

$$y_{ij} = \langle S_i^x S_j^x \rangle - \langle S_i^y S_j^y \rangle, \tag{6}$$

where $\langle S_j^z \rangle$ is the magnetization density at site $j$,

$$\langle S_j^z \rangle = -\frac{1}{N} \sum_{p>0}^{N/2} \frac{\tanh(\beta \Delta / 2)}{\Delta} \left[ J(\cos\phi_p - 2D\sin\phi_p) - 1 \right], \tag{7}$$

where $\Delta = \sqrt{\left[ J(\cos\phi_p - 2D\sin\phi_p) - 1 \right]^2 + J^2 \gamma^2 \sin^2 \phi_p}$ with $\phi_p = 2\pi p / N$, $p = -N/2, ..., N/2$, and $\beta = 1/kT$ with $k$ being Boltzmann constant and $T$ the absolute temperature. When the system has translation invariant, we obtain $\langle S_i^z \rangle = \langle S_j^z \rangle$ ($\forall i, j$) such that $\omega_{ij}^+ = \omega_{ij}^-$. By using the method in Ref. [41], we can find analytically the two-point correlation functions $\langle S_i^\alpha S_j^\beta \rangle$ ($\alpha, \beta = x, y, z$) at the subsystems $i$ and $j$ [42],

$$\langle S_i^x S_j^x \rangle = \frac{1}{4} \begin{vmatrix} G_{-1} & G_{-2} & \cdots & G_{i-j} \\ G_0 & G_{-1} & \cdots & G_{i-j+1} \\ \cdots & \cdots & \cdots & \cdots \\ G_{j-i-2} & G_{j-i-3} & \cdots & G_{-1} \end{vmatrix}, \tag{8}$$

$$\langle S_i^y S_j^y \rangle = \frac{1}{4} \begin{vmatrix} G_1 & G_0 & \cdots & G_{i-j+2} \\ G_2 & G_1 & \cdots & G_{i-j+3} \\ \cdots & \cdots & \cdots & \cdots \\ G_{j-i} & G_{j-i-1} & \cdots & G_1 \end{vmatrix}, \tag{9}$$

and

$$\langle S_i^z S_j^z \rangle = \langle S^z \rangle^2 - \frac{1}{4} G_{j-i} G_{i-j}, \tag{10}$$

with

$$G_R = -\frac{1}{N} \sum_{p>0}^{N/2} 2\cos(\phi_p R) \left[ J(\cos\phi_p - 2D\sin\phi_p) - 1 \right] \frac{\tanh(\beta \Delta / 2)}{\Delta}$$

$$+\frac{\gamma}{N}\sum_{p>0}^{N/2} 2J \sin(\phi_p R)\sin\phi_p \frac{\tanh(\beta\Delta/2)}{\Delta}. \tag{11}$$

When $N \to \infty$, the limit values can be obtained by replacing $\phi_p$ by $\phi$ and the sum by integral $\frac{1}{2\pi}\int_0^\pi d\phi$. The eigenvalues of the density matrix $\rho_{ij}$ are

$$\lambda_1 = \frac{1}{4}\left(1 - 4\langle S_i^x S_j^x\rangle - 4\langle S_i^y S_j^y\rangle - 4\langle S_i^z S_j^z\rangle\right), \tag{12}$$

$$\lambda_2 = \frac{1}{4}\left(1 + 4\langle S_i^x S_j^x\rangle + 4\langle S_i^y S_j^y\rangle - 4\langle S_i^z S_j^z\rangle\right), \tag{13}$$

$$\lambda_3 = \frac{1}{4}\left[1 - 4\sqrt{\left(\langle S_i^x S_j^x\rangle - \langle S_i^y S_j^y\rangle\right)^2 + \langle S^z\rangle^2} + 4\langle S_i^z S_j^z\rangle\right], \tag{14}$$

$$\lambda_4 = \frac{1}{4}\left[1 + 4\sqrt{\left(\langle S_i^x S_j^x\rangle - \langle S_i^y S_j^y\rangle\right)^2 + \langle S^z\rangle^2} + 4\langle S_i^z S_j^z\rangle\right]. \tag{15}$$

We now come to briefly review the definition of the pairwise quantum discord. A bipartite quantum state $\rho_{ij}$ contains both classical and quantum correlations, and the total correlations are measured jointly by the quantum mutual information. The total correlations between subsystems $i$ and $j$ are

$$\mathcal{I}(\rho_{ij}) = S(\rho_i) + S(\rho_j) - S(\rho_{ij}), \tag{16}$$

where $\rho_{i(j)} = \mathrm{Tr}_{j(i)}(\rho_{ij})$ is the reduced density matrix of the subsystem $i(j)$, and $S(\rho) = \mathrm{Tr}\rho \log \rho$ is the von Neumann entropy, and $\log$ represents the logarithm with base 2. The quantum discord is defined as the difference between the total correlations and the classical correlation,

$$\mathcal{QD}(\rho_{ij}) = \mathcal{I}(\rho_{ij}) - \mathcal{CC}(\rho_{ij}), \tag{17}$$

where the classical correlation $\mathcal{CC}(\rho_{ij})$ between the subsystems is given by [43]

$$\mathcal{CC}(\rho_{ij}) = S(\rho_i) - \min_{\{P_\kappa\}}\left[p_\kappa S(\rho_i^\kappa)\right], \tag{18}$$

where $\{P_\kappa\}$ is a set of projects performed locally on the subsystem $j$, and $\rho_i^\kappa = \mathrm{Tr}_j\left[(I_i \otimes P_\kappa)\rho_{ij}(I_i \otimes P_\kappa)\right]/p_\kappa$ is the quantum state of the subsystem $i$ conditioned on the measurement outcome labeled by $\kappa$, with probability $p_\kappa = \mathrm{Tr}\left[(I_i \otimes P_\kappa)\rho_{ij}(I_i \otimes P_\kappa)\right]$.

Here $I_i$ is the identity operator for the subsystem $i$. The mutual information (16) reads

$$\mathcal{I}(\rho_i : \rho_j) = 2\left(-\eta_1 \log \eta_1 - \eta_2 \log \eta_2\right) + \sum_{\upsilon=1}^{4} \lambda_\upsilon \log \lambda_\upsilon, \tag{19}$$

where $\eta_{1,2} = 1/2 \pm \langle S_i^z \rangle$ and $\lambda_\upsilon$ ($\upsilon = 1, 2, 3, 4$) correspond to eigenvalues of the reduced density matrices $\rho_i$ and $\rho_{ij}$, respectively. The condition $S(\rho_i) = S(\rho_j)$ is satisfied such that the measurement of classical correlation assumes equal values, irrespective of whether the measurement is performed on the site $i$ or $j$. Here we consider the complete set of orthogonal projectors $\{P_\kappa(\theta, \phi) = I \otimes |\kappa\rangle\langle\kappa|\}$ ($\kappa = 1, 2$) for a local measurement performed on the site $j$, where the two projectors are defined by the spin states

$$|1\rangle = \cos\theta |\uparrow\rangle + e^{i\phi} \sin\theta |\uparrow\rangle, \tag{20}$$

$$|2\rangle = \sin\theta |\uparrow\rangle - e^{i\phi} \cos\theta |\downarrow\rangle. \tag{21}$$

It is straightforward to calculate the quantum discord [19]

$$\mathcal{QD}(\rho_{ij}) = \min\{\mathcal{QD}_1, \mathcal{QD}_2\}, \tag{22}$$

where

$$\mathcal{QD}_1 = S(\rho_i) - S(\rho_{ij}) - u_{ij}^+ \log\left(\frac{u_{ij}^+}{u_{ij}^+ + \omega_{ij}}\right) - \omega_{ij} \log\left(\frac{\omega_{ij}}{u_{ij}^+ + \omega_{ij}}\right)$$
$$- u_{ij}^- \log\left(\frac{u_{ij}^-}{u_{ij}^- + \omega_{ij}}\right) - \omega_{ij} \log\left(\frac{\omega_{ij}}{u_{ij}^- + \omega_{ij}}\right), \tag{23}$$

and

$$\mathcal{QD}_2 = S(\rho_i) - S(\rho_{ij}) - \frac{1+\Lambda}{2} \log\frac{1+\Lambda}{2} - \frac{1-\Lambda}{2} \log\frac{1-\Lambda}{2}, \tag{24}$$

with $\Lambda = \sqrt{\left(u_{ij}^+ - u_{ij}^-\right)^2 + 4\left(|x_{ij}| + |y_{ij}|\right)^2}$.

The quantum discord is a measure of nonclassical correlations. It may include but is independent of entanglement. For pure states and a mixture of Bell states, the quantum discord is equivalent to the entanglement entropy. However, for general two-qubit mixed states, the correspondence between them is much complicated. The following work will focus on the quantum discord, entanglement, and the classical correlation in the anisotropic $XY$ spin-half chain with the DM interaction. We use concurrence as the measure of entanglement [44], defined by

$$C(\rho_{ij}) = \max\{\varsigma_1 - \varsigma_2 - \varsigma_3 - \varsigma_4, 0\}, \qquad (25)$$

where $\varsigma_\mu$ ($\mu = 1, 2, 3, 4$) are the square roots of the eigenvalues in descending order of the operator $R_{ij} = \rho_{ij}\tilde{\rho}_{ij}$, $\tilde{\rho}_{ij} = \left(\sigma_i^y \otimes \sigma_j^y\right)\rho_{ij}^*\left(\sigma_i^y \otimes \sigma_j^y\right)$, with $\rho_{ij}^*$ being the conjugate of $\rho_{ij}$, and $\sigma_{i(j)}^y$ is the $y$ component of the Pauli matrix for the site $i$ or $j$. The concurrence of the density matrix (2) is

$$C(\rho_{ij}) = 2\max\{0, |x_{ij}| - \sqrt{u_{ij}^+ u_{ij}^-}, |y_{ij}| - \omega_{ij}\}. \qquad (26)$$

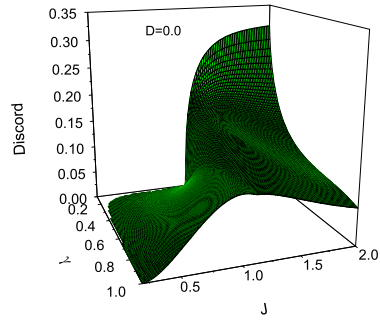

(a)

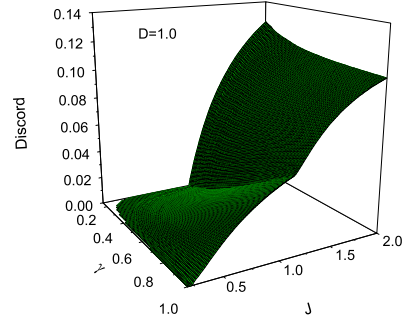

(b)

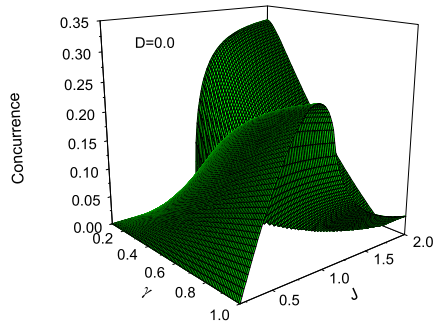

(c)

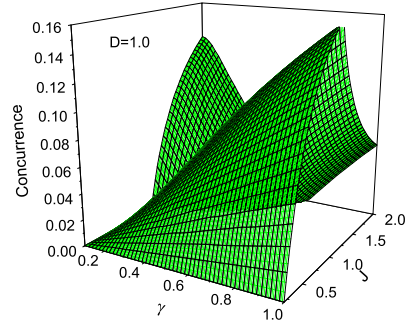

(d)

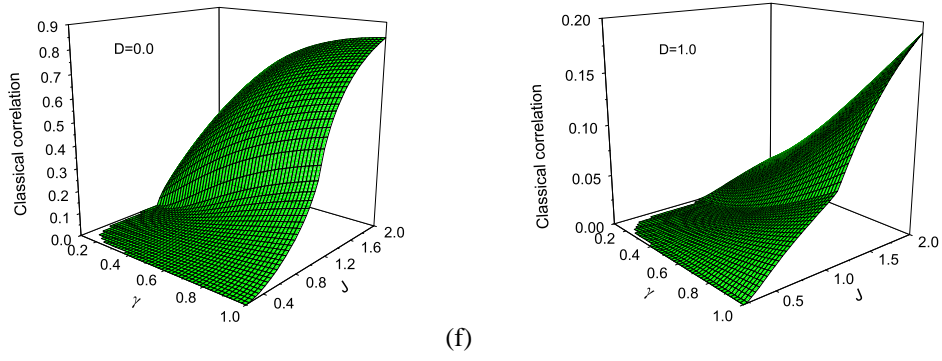

(e)                                                                    (f)

FIG.1. (Color online) Quantum discord (a) (b), entanglement (c) (d), and classical correlation (e) (f) for the nearest-neighbor spins in the $XY$ spin chain as a function of the anisotropy $\gamma$ and magnetic intensity $J$ at zero temperature for different values of DM interaction.

In the thermodynamic limit, we plot the pairwise quantum discord, entanglement, and classical correlation in the $XY$ chain for different values of DM interaction in Figs.1(a)-1(f), respectively. Fig. 1(a) and Fig. 1(b) show that there is a clear difference in quantum discord between the two regions $J \in [0,1]$ and $J \in [1,2]$. The quantum phase transition occurs at the critical point $J=1$. When $J \leq 1$, the quantum discord increases with anisotropy and reaches the maximum at the point $\gamma = 1$. When $J > 1$, the quantum discord decays monotonously with the anisotropy.

We now turn our attention to the effect of the DM interaction on quantum discord. It seems that the quantum discord is suppressed by the DM interaction. Furthermore, in the absence of the anisotropy, the quantum discord is insensitive to the DM interaction.

Before exploring the numerical details, it is interesting to note that there is a family of Hamiltonians containing the DM interaction, leading to the same entanglements and discord [34]. In this family, the Hamiltonians are subject to single spin unitary transformations,

$$W = \prod_{j=1}^{N} \exp\left[-i\left(\alpha_j S_j^x + \beta_j S_j^y + \delta_j S_j^z\right)\right], \tag{27}$$

such that $H' = WHW^\dagger$, where $\alpha_j$, $\beta_j$ and $\delta_j$ are independent angles. For instance, if $W = \prod_{j=2,4,6,\ldots} \exp\left[-i\pi/2 S_j^z\right]$ where $j$s are even numbers, the Hamiltonian (1) can be rewritten as

$$H' = \sum_{j=1}^{N} \left\{ J\left[ -(1+\gamma) S_j^x S_{j+1}^y + (1-\gamma) S_j^y S_{j+1}^x + D\left( S_j^x S_{j+1}^x + S_j^y S_{j+1}^y \right) \right] - S_j^z \right\}. \tag{28}$$

The DM interaction plays a special role in the isotropic $XY$ model with $\gamma = 0$. Based on the following relations,

$$\exp(-i\vartheta S_j^z) S_j^x \exp(i\vartheta S_j^z) = S_j^x \cos\vartheta + S_j^y \sin\vartheta, \tag{29}$$

$$\exp(-i\vartheta S_j^z) S_j^y \exp(i\vartheta S_j^z) = -S_j^x \sin\vartheta + S_j^y \cos\vartheta, \tag{30}$$

the transformation $W = \prod_{j=2,4,6,\ldots} \exp\left[i\vartheta S_j^z\right]$ transfers the isotropic $XY$ chain into

$$H' = \sum_{j=1}^{N} \left\{ J'\left[ S_j^x S_{j+1}^x + S_j^y S_{j+1}^y + D'\left( S_j^x S_{j+1}^y - S_j^y S_{j+1}^x \right) \right] - S_j^z \right\}, \tag{31}$$

where $J' = J\cos\vartheta$, and $D' = \tan\vartheta$. It implies that the DM force in the isotropic case is given by single spin transformation and does not make a significant contribution to the entanglement and quantum discord.

The pairwise entanglement for the general Hamiltonian (1) is displayed in Fig. 1(c) and Fig. 1(d). As shown in the figures, the influence of the anisotropy $\gamma$ on the entanglement is different between the $J < 1$ region and the $J > 1$ region. The DM interaction in the $XY$ spin chain suppresses the pairwise entanglement and the pairwise quantum discord. In contrast to the quantum discord and entanglement, the classical correlation always increases with the anisotropy as shown in the last subfigures of Fig. 1.

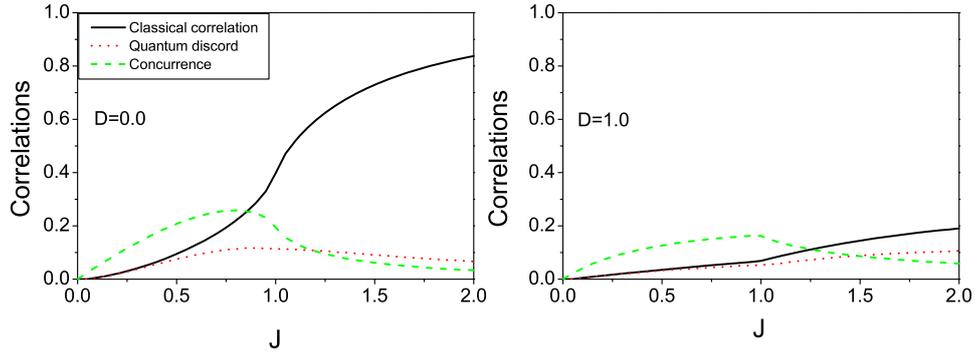

FIG.2. (Color online) Quantum discord (dotted line), entanglement (dashed line), and classical correlation (solid line) as a function of $J$, for the nearest-neighbor spins in the transverse Ising chain, (a) $D = 0.0$, (b) $D = 1.0$.

Fig. 2 displays the quantum discord (dotted line), entanglement (dashed line), and classical correlation (solid line) as a function of $J$ for the nearest-neighbor spins in the transverse Ising model ($\gamma = 1$) with different values of the DM interaction. It is clear that, in this particular case, both classical and quantum correlations are suppressed by the DM interaction. While the concurrence is initially larger than the quantum discord and classical correlation and then becomes less, the quantum discord is always less than classical correlation. Therefore it is reasonable to conclude that the quantum discord is a measure independent of entanglement, and the three correlations are substantially different qualitatively and quantitatively, as also noticed in a recent paper [12].

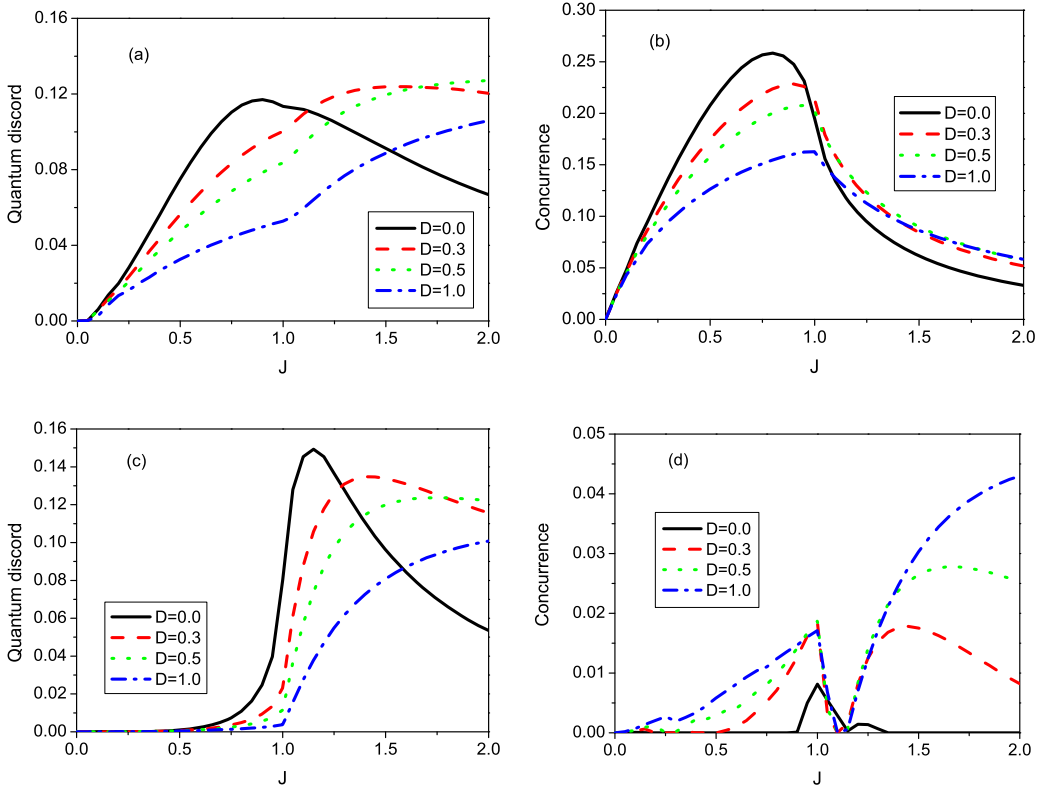

FIG.3. (Color online) Quantum discord (a) and entanglement (b) for the nearest-neighbor spins as a function of $J$ for different DM interaction, when $\gamma = 1.0$. (c) and (d) show cases of the pairwise quantum correlations for the third nearest neighbor spins, when $\gamma = 0.5$.

In order to better understand the DM interaction, we plot pairwise quantum discord and entanglement against $J$, with different DM interaction in Fig.3. The pairwise quantum discord shows distinct behaviors when the parameter D varies. In the region $J \in (0,1]$, the quantum discord can be suppressed by the DM interaction. The larger the DM interaction is, the less the quantum discord is. In the region $J \in [1,2]$, the pairwise quantum discord without the DM interaction decays monotonously. However, when the DM interaction is present, the situation becomes more complicated as shown in this figure.

The entanglement for nearest-neighbor spins is presented in Fig.3 (b). The concurrence always has maximum around the critical point $J=1$, regardless of the DM interaction. When $J<1$ the DM interaction restraints the growth of entanglement, while it rarely changes the decay rate of entanglement in the $J>1$ region.

The quantum correlations decrease expectedly as the site distance increases. We plot the quantum discord and entanglement for third nearest spins with the anisotropy parameter $\gamma=0.5$ in Figs. 3 (c)-(d). The pairwise entanglement is tiny ($C_{max}\approx 0.04$) or disappears completely for most of values of $\gamma$. On the contrary, the quantum discord remains nonzero even for farther site distances. This may make quantum discord more promising in many aspects of quantum physics, such as being an indicator to a quantum phase transition [24]. Moreover, we observe that without the DM interaction the entanglement is zero expect in the vicinity of the critical point, as depicted in Fig. 3 (d). It is interesting to note that the DM interaction enhances the concurrence greatly, which may imply that the DM interaction is able to promote the long-distance entanglement.

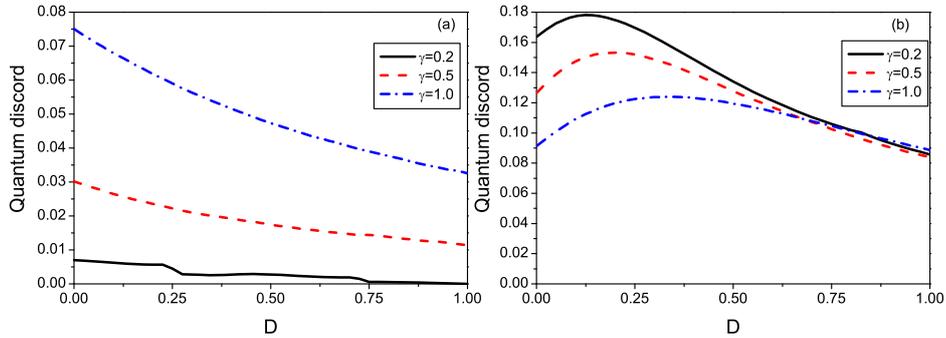

FIG.4. (Color online) The quantum discord for the nearest-neighbor spins in the $XY$ spin chain as a function of the DM interaction with different values of the anisotropy. (a) $J=0.5$. (b) $J=1.5$.

Fig.4 plots the quantum discord against the DM interaction for different values of $J$ and anisotropy $\gamma$. When $J<1$, the quantum discord as a function of D decays monotonously. When $J>1$, curves of quantum discord have peaks at $D \approx 0.25$. This supports again that the quantum discord is enhanced in the region $J \in (0,1]$ but be reduced in the $J > 1$ region as the anisotropy increases.

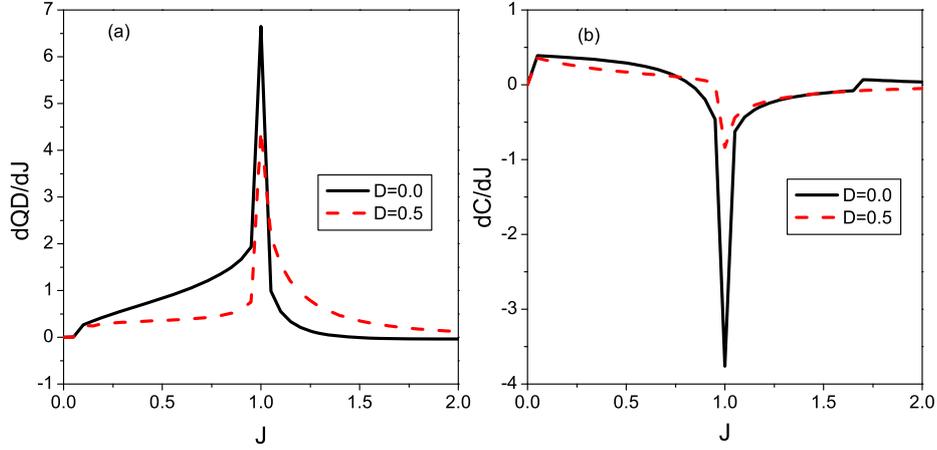

FIG.5. (Color online) First derivative of the quantum discord and entanglement for the nearest-neighbor spins with respect to $J$ for different values of DM interaction with $\gamma = 0.8$.

There exists a second order quantum phase transition at $J = 1$ in the model (1). We plot the first derivative of the nearest-neighbor quantum discord and entanglement with respect to $J$ for different values of the DM interaction in Fig.5. The non-analyticities of $dQD/dJ$ and $dC/dJ$ expectedly appear at the critical point $J = 1$ for any value of the DM interaction, which signals the quantum phase transition. The first derivative of classical correlation $d\mathcal{CC}/dJ$ can also signal the quantum phase transition. Both $dQD/dJ$ and $d\mathcal{CC}/dJ$ have a pronounced maximum at the critical point, with the same behavior as $dC/dD$ [23]. However, it should be emphasized that the DM interaction weakens the critical behavior in the derivatives of these correlations.

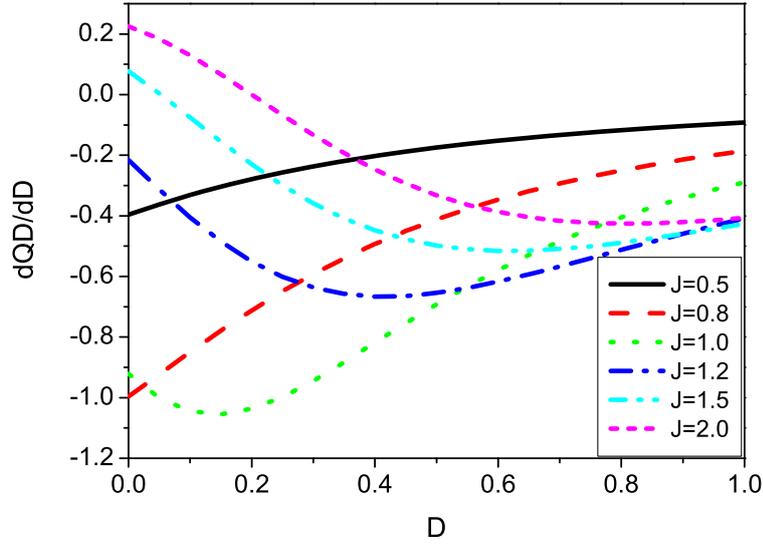

FIG.6. (Color online) First derivative of the quantum discord with respect to the DM interaction for different values of $J$ with $\gamma =1$.

Although the DM interaction does not change the universality class of quantum phase transition of the present model, we plot $dQD/dD$ as a function of the DM interaction as an analogy in Fig.6. While no singularity can be observed as expected, the derivative of quantum discord with respect to the DM interaction $dQD/dD$ shows distinct behaviors for different regions of $J$.

## IV. CONCLUSIONS

We have studied the quantum discord, entanglement, and classical correlation for the nearest-neighbor spins in the anisotropy $XY$ spin-half chain with DM interaction. The quantum correlations increase monotonously with the anisotropy in the $J<1$ region, while in the $J>1$ region the quantum correlations may decrease as the anisotropy increases. On the contrary, the pairwise classical correlation always increases with the anisotropy. The DM interaction is the central topic of this paper. Our analysis shows that while the DM interaction suppresses the standard behaviors of the anisotropic XY model, it enhances surprisingly the values of long-distance correlations greatly. However, when the anisotropy parameter is very small, the quantum correlation is not significantly affected by the DM interaction, as shown generally in our family of Hamiltonians.

The role of the DM interaction can also be understood by analyzing the derivatives of both quantum and classical correlations with respect to $J$ and the DM interaction. While the singularities in quantum discord $dQD/dJ$, entanglement $dC/dJ$, and classical correlation $d\mathcal{CC}/dJ$ at the critical point $J=1$ indicate the quantum phase transition, they are weaken and smoothen by the DM interaction. In addition, no nonanalyticity in $dQD/dD$ indicates, as well expected, that the universality class of the model is not affected by the DM interaction.

## ACKNOLEDGMENTS

We acknowledge financial support by National Natural Science Foundation of China under Grant Nos. 11075013 and 10974016, L. –A. Wu has been supported by the Ikerbasque Foundation Start-up, the Basque Government (grant IT472-10) and the Spanish MEC (Project No. FIS2009-12773-C02-02). We thank Drs. Ping Lou and M. S. Sarandy for helpful suggestions.

## REFERENCES

[1] M. A. Nielsen and I. L. Chuang, *Quantum Computation and Quantum Information* (Cambridge University Press, Cambridge, UK, 2000).

[2] L. Amico, R. Fazio, A. Osterloh, and V. Vedral, Rev. Mod. Phys. **80**, 517 (2008).

[3] L. Henderson and V. Vedral, J. Phys. A **34**, 6899 (2001).

[4] H. Ollivier and W. H. Zurek, Phys. Rev. Lett. **88**, 017901 (2001).

[5] B. Groisman, S. Popescu, and A. Winter, Phys. Rev. A **72**, 032317 (2005).

[6] K. Modi, T. Paterek, W. Son, V. Vedral, and M. Williamson, Phys. Rev. Lett. **104**, 080501 (2010).

[7] S. Luo, Phys. Rev. A **77**, 042303 (2008); A. Datta, *ibid.* **80**, 052304 (2009).

[8] R. Vasile, P. Giorda, S. Olivares, M. G. A. Paris, and S. Maniscalco, Phys. Rev. A **82**, 012313 (2010).

[9] A. Ferraro, L. Aolita, D. Cavalcanti, F. M. Cucchietti, and A. Acín, Phys. Rev. A **81**, 052318 (2010).

[10] L. Mazzola, J. Piilo, and S. Maniscalco, Phys. Rev. Lett. **104**, 200401 (2010).

[11] R. –C. Ge, M. Gong, C. –F. Li, J.-S. Xu, and G.-C. Guo, Phys. Rev. A **81**, 064103 (2010); J. –B. Yuan, J. –Q. Liao, and L. –M. Kuang, arXiv:1005.4204.

[12] M. Ali, A. R. P. Rau, and G. Alber, Phys. Rev. A **81**, 042105 (2010).

[13] L. Ciliberti, R. Rossignoli, and N. Canosa, Phys. Rev. A **82**, 042316 (2010).

[14] A. Datta, A. Shaji, and C. M. Caves, Phys. Rev. Lett. **100**, 050502 (2008); A. Datta, Phys. Rev. A **80**, 052304 (2009).

[15] B. P. Lanyon, M. Barbieri, M. P. Almeida, and A. G. White, Phys. Rev. Lett. **101**, 200501 (2008).

[16] J. Cui and H. Fan, J. Phys. A: Math. Theor. **43**, 045305 (2010).

[17] T. Werlang, S. Souza, F. F. Fanchini, and C. J. Villas Boas, Phys. Rev. A **80**, 024103 (2009); T. Werlang and Gustavo Rigolin, *ibid.* **81**, 044101(2010).

[18] B. Wang, Z.-Y. Xu, Z. –Q. Chen, and M. Feng, Phys. Rev. A **81**, 014101 (2010).

[19] F. F. Fanchini, T. Werlang, C. A. Brasil, L. G. E. Arruda, and A. O. Caldeira, Phys. Rev. A **81**, 052107 (2010).

[20] G. Adesso and A. Datta, Phys. Rev. Lett. **105**, 030501 (2010).

[21] P. Giorda, and M. G. A. Paris, Phys. Rev. Lett. **105**, 020503 (2010).

[22] R. Dillenschneider, Phys. Rev. B **78**, 224413 (2008); M. S. Sarandy, Phys. Rev. A **80**, 022108 (2009).

[23] T. Werlang, C. Trippe, G. A. P. Ribeiro, and Gustavo Rigolin, Phys. Rev. Lett. **105**, 095702 (2010).

[24] J. Maziero, H. C. Guzman, L. C. Céleri, M. S. Sarandy, and R. M. Serra, Phys. Rev. A **82**, 012106 (2010). L.-A. Wu, M.S. Sarandy, and D.A. Lidar, Phys. Rev. Lett. Phys. Rev. Lett. **93**, 250404 (2004).

[25] I. Dzyaloshinsky, J. Phys. Chem. Solids **4**, 241(1958); T. Moriya, Phys. Rev. Lett. **4**, 228 (1960).

[26] G. F. Zhang, Phys. Rev. A **75**, 034304 (2007).

[27] M. K. Kwan, Z. N. Gurkan, and L. C. Kwek, Phys. Rev. A **77**, 062311 (2008).

[28] D. C. Dender, P. R. Hammar, D. H. Reich, C. Broholm, and G. Aeppli, Phys. Rev. Lett. **79**, 1750 (1997).

[29] M. Kohgi, K. Iwasa, J. M. Mignot, B. Fak, P. Gegenwart, M. Lang, A. Ochiai, H. Aoki, and T. Suzuki, Phys. Rev. Lett. **86**, 2439 (2001).

[30] I. Tsukada, J. T. Takeya, T. Masuda, and K. Uchinokura, Phys. REV. Lett. **87**, 127203 (2001).

[31] M. Greven, R. J. Birgeneau, Y. Endoh, M. A. Kastner, M. Matsuda, and G. Shirane, Z. Phys. B: Condens. Matter 96, 465 (1995).

[32] K. V. Kavokin, Phys. Rev. B **64**, 075305(2001).

[33] L. –A. Wu and D. A. Lidar, Phys. Rev. A **66**, 062314 (2002).

[34] L. –A. Wu and D. A. Lidar, Phys. Rev. Lett. **91**, 097904 (2003)

[35] R. Jafari, M. Kargarian, A. Langari, and M. Siahatgar, Phys. Rev. B **78**, 214414 (2008); M. Kargarian, R. Jafari, and A. Langari, Phys. Rev. A **79**, 042319 (2009).

[36] B. Wang, M. Feng, and Z.-Q. Chen, Phys. Rev. A **81**, 064301 (2010).

[37] W. Wernsdorfer, T. C. Stamatatos, and G. Christou, Phys. Rev. Lett. **101**, 237204(2008).

[38] K. Maruyama, T. Iitaka, and F. Nori, Phys. Rev. A **75**, 012325 (2007).

[39] M. Rafiee, M. Soltani, H. Mohammadi, and H. Mokhtari, arXiv:1010.0387.

[40] J.-M. Cai, Z.-W. Zhou, G. –C. Guo, Phys. Lett. A **352**, 196 (2006); X. Wang, P. Zanardi, *ibid.* **301**, 1 (2002).

[41] E. Lieb, T. Schultz, and D. Mattis, Ann. Phys. (N. Y.) **16**, 407 (1961).

[42] E. Barouch, B. M. McCoy, Phys. Rev. A **3**, 786 (1971).

[43] W. H. Zurek, Rev. Mod. Phys. **75**, 715(2003); L. Henderson and V. Vedral, J. Phys. A **34**, 6899(2001); V. Vedral, Phys. Rev. Lett. **90**, 050401(2003).

[44] W. K. Wootters, Phys. Rev. Lett. **80**, 2245(1998).